\documentstyle[floats,aps,epsf,prl]{revtex}

\begin{document}
\draft
\twocolumn[\hsize\textwidth\columnwidth\hsize\csname @twocolumnfalse\endcsname

\title{Quantum Hall effect at low magnetic fields}

\author{Bodo Huckestein\cite{p_address}}

\address{Institut f\"ur Theoretische Physik, Universit\"at zu
  K\"oln, D-50937 K\"oln, Germany}

\date{30 June 1999}

\maketitle

\begin{abstract}
  The temperature and scale dependence of resistivities in the
  standard scaling theory of the integer quantum Hall effect is
  discussed. It is shown that recent experiments, claiming to observe
  a discrepancy with the global phase diagram of the quantum Hall
  effect, are in fact in agreement with the standard theory. The
  apparent low-field transition observed in the experiments is
  identified as a crossover due to weak localization and a strong
  reduction of the conductivity when Landau quantization becomes
  dominant.
\end{abstract}
\vskip2pc

]

The behavior of the quantum Hall effect (QHE) at low magnetic fields
has attracted a lot of attention in recent years, both experimentally
and theoretically. Of particular interest has been the fate of the
critical states responsible for the transitions between the different
quantized Hall plateaus and the form of the global phase diagram of
the QHE \cite{KLZ92,Huc95r}. Recent experiments
\cite{SKD93,KMFP95,Sea97,LCSL98,Hea99} have been interpreted as being
incompatible with the global phase diagram \cite{KLZ92} based on the
levitation of critical states \cite{Khm84,Lau84}. At high magnetic
fields, such that the Landau level separation is much larger than the
disorder broadening of the Landau bands, these critical states are
situated near the centers of the Landau bands. They separate phases
with different, quantized values of the Hall conductivity. As the Hall
conductivity is constant throughout these phases, the critical
energies cannot just terminate when magnetic field or disorder is
changed. They either can move to infinite energy or they can terminate
when they intersect another critical state with the opposite change in
Hall conductivity.  Both of these scenarios are realized: the critical
energies at the centers of the Landau bands move to infinite energy as
the magnetic field becomes infinitely strong, while the termination of
critical states has been observed in lattice models, where due to the
band structure, states with negative Hall conductivity exist
\cite{LXN96}.

\begin{figure}
  \begin{center}
    \epsfxsize=6.5cm
    \leavevmode
    \epsffile{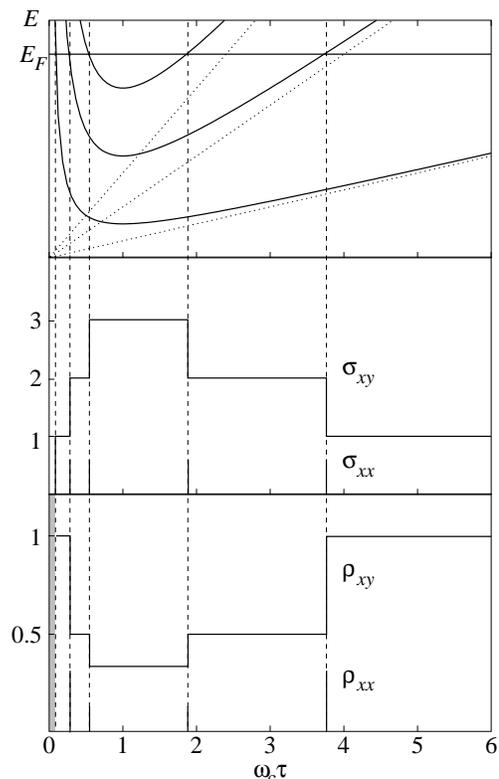}
    \caption[oskar]{Sketch of the magnetic field dependence of the critical
      energies according to the levitation picture
      ($E_n=(n+1/2)\hbar\omega_c(1+1/(\omega_c\tau)^2)$ \cite{Lau84})
      (top), the corresponding (dimensionless) conductivities
      (center), and resistivities (bottom).}
    \label{fig:scale}
  \end{center}
\end{figure}

Since in the absence of a periodic potential, no states with negative
Hall conductivity exist, Khmel'nitskii and Laughlin have argued, that
the existence of critical states at high magnetic fields can only be
reconciled with the absence of extended states at zero magnetic field,
predicted by the scaling theory of localization in two dimensions
\cite{AALR79}, if the critical states float above the Fermi energy,
when the magnetic field is decreased towards zero. The scaling theory
of the integer QHE \cite{scaling} predicts the Hall conductivity to be
quantized at $ne^2/h$ between the critical energies, with integer $n$,
and to be $(n+1/2)e^2/h$ at the critical energies\cite{Huc95r}. The
dissipative conductivity vanishes, except at the critical energies,
where it takes on values of the order of $e^2/h$. In
Fig.~(\ref{fig:scale}) the levitation of the critical states is
sketched and the resulting behavior of the conductivities and
resistivities is shown. Note, that it is not possible to predict the
behavior of the resistivities in the phase with zero Hall
conductivity, as their values depend on the way how the conductivities
tend to zero.

Since the QH plateau transitions are quantum phase transitions, the
discussion presented above is concerned with the phase diagram of an
infinite system at zero temperature. Experimentally, only finite
systems at finite temperature are accessible. In numerical
simulations, also the restriction to finite systems applies. When
comparing the result of experiments with the predictions of scaling
theory, it is therefore imperative to consider the effects of finite
temperatures and/or finite system sizes. While this was appreciated in
early theoretical and experimental work \cite{Khm84,WTP85}, it has
wandered out of the focus of much of the recent work. It is the
purpose of this paper to show that the experiments on the low-field
QHE can be understood within the standard scaling theory, obviating
the need for a more exotic explanation. We will restrict our attention
to the integer QHE and consider interaction effects only on the level
of weak localization corrections.

\begin{figure}
  \begin{center}
    \epsfxsize=6.5cm
    \leavevmode
    \epsffile{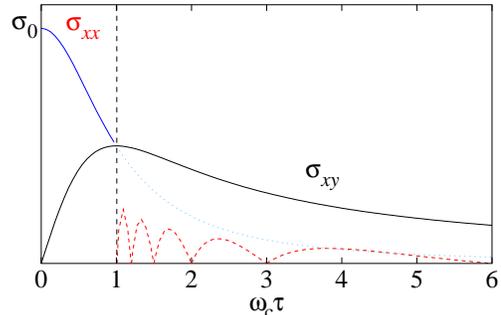}
    \caption{Conductivities on short length scales: the Drude results
      (solid lines and dotted line) and the SCBA result for
      $\sigma_{xx}$ (dashed), appropriate for $\omega_c\tau\gg1$.}
    \label{fig:drude}
  \end{center}
\end{figure}

We will start our argument by considering the appropriate starting
points for a renormalization of the conductivities, the bare
conductivities $\sigma_{ij}^0$, corresponding to short length scales
or high temperatures \cite{greenwood}. At high temperatures and low
magnetic fields, quantum effects are negligible and the conductivities
can be calculated from kinetic equations to give the Drude expressions
\begin{mathletters}
  \label{eq:drude}
  \begin{eqnarray}
    \sigma_{xx}^0 &=& \frac{\sigma_0}{1 + (\omega_c \tau)^2},\\
    \sigma_{xy}^0 &=& \omega_c\tau\sigma_{xx}^0,
  \end{eqnarray}
\end{mathletters}%
with $\sigma_0=e^2 n_c \tau/m^*=en_c\mu$, $\omega_c=eB/m^*$, and
$n_c$, $\tau=\ell/v_F$, $\ell$, and $\mu$ are the carrier density,
transport time, elastic mean free path, and the mobility, respectively
(Fig.~(\ref{fig:drude})). In terms of resistivities, the classical
values are $\rho_{xx}^0=1/\sigma_0$, independent of the magnetic field
$B$, and $\rho_{xy}^0=B/en$. Quantum effects modify these results in
two ways: quantum interference leads to localization, and Landau
quantization drastically modifies the density of states at strong
magnetic fields. Quantum mechanically, three energy scales are
relevant: the cyclotron energy $E_B=\hbar\omega_c$, the disorder
broadening of the Landau bands $\Gamma$, and the thermal energy
$E_T=k_BT$. The ratio of the former two depends on the strength of the
magnetic field and the strength and range of the disorder \cite{AU74}
and corresponds to the classical quantity $\omega_c\tau$,
characterizing the classical effects of the magnetic field and
disorder in eqs.~(\ref{eq:drude}). While $E_B/\Gamma$ and
$\omega_c\tau$ are not identical, $\omega_c\tau=1$ can serve as an
estimate of the point where Landau level quantization becomes
important. The ratio of cyclotron energy to temperature,
$E_B/E_T=\hbar e B/m^* k_B T$, determines whether the classical
expression for $\sigma_{xx}$ is appropriate or Landau quantization has
to be taken into account.  In the limit of strong magnetic fields,
such that Landau level mixing can be neglected, the high temperature
(or short length scale) conductivity is qualitatively well described
within the self-consistent Born approximation (SCBA) \cite{AU74}. In
this approximation, the conductivity vanishes at zero temperature for
integer filling factors $\nu=n_c2\pi l_c^2$, with the magnetic length
$l_c^2=\hbar/eB$, due to a vanishing density of states. For smooth
random potential, relevant to most experiments on GaAs/AlGaAs
heterostructures, the peak value of $\sigma_{xx}^0$ in SCBA is given
by $(l_c^2/\pi d^2)(e^2/h)$, independent of the Landau level index.
$d$ is the range of the disorder potential
(Fig.~(\ref{fig:drude}))\cite{scba2}. Landau level quantization can
thus lead to a strong reduction in the conductivity compared to the
Drude result, provided that the magnetic field is strong enough, i.e.
$\omega_c\tau>1$.

The localizing effect of quantum interference is most important for
the occurrence of the QHE and it is the key to our understanding of the
temperature dependence of the resistivities. In the absence of a
magnetic field, quantum interference leads to a size-dependent
reduction of the conductivity \cite{LR85}
\begin{equation}
  \label{eq:xx_cor}
  \sigma_{xx}(L) = \sigma_{xx}^0 - \frac{2e^2}{\pi h} \log
  \left(\frac{L}{\ell}\right).
\end{equation}
This weak localization expression is valid for large $\sigma_{xx}$.
While the system size dependence is logarithmically weak for small
system sizes, scaling theory predicts that it will eventually lead to
complete localization and vanishing conductivity. The corresponding
corrections to the Hall conductivity are given by \cite{Fuk80}
\begin{equation}
  \label{eq:xy_cor}
  \sigma_{xy}(L) = \sigma_{xy}^0 - \omega_c\tau \frac{4e^2}{\pi h} \log
  \left(\frac{L}{\ell}\right).
\end{equation}
Again, the decrease in the Hall conductivity is the precursor of the
vanishing Hall conductivity at low fields predicted by scaling theory
(Fig.~(\ref{fig:scale})).  In terms of the resistivities, these
corrections lead to a logarithmic increase in the dissipative
resistivity, while the Hall resistivity remains unchanged. In addition
to the disorder effects, Coulomb interactions lead to logarithmic
corrections to $\sigma_{xx}$, but not to $\sigma_{xy}$ \cite{LR85}.
While these effects are important for a detailed comparison with
experiment, they do not change the conclusions of the present
discussion and will be neglected in the following.

In the presence of a magnetic field, quantum interference effects are
reduced and the system size dependence of $\sigma_{xx}$ becomes even
weaker \cite{Hik81Efe83},
\begin{equation}
  \label{eq:uni_cor}
  \sigma_{xx}(L) = \sigma_{xx}^0 -
  \frac{1}{\pi^2\sigma_{xx}^0}\left(\frac{e^2}{h}\right)^2\log
  \left(\frac{L}{l_c}\right).
\end{equation}
This means that localization effects become strong, when the system
size exceeds the localization length
\begin{equation}
  \label{eq:xi0}
  \xi^0=l_c \exp(\pi^2\sigma_{xx}^0{}^2h^2/e^4),
\end{equation}
defined by $\sigma_{xx}(\xi^0)=0$. In contrast to the zero field case,
the system then does not become completely localized but exhibits a
series of critical energies at which the conductivity remains finite
and the Hall conductivity changes by $e^2/h$ as shown in
Fig.~(\ref{fig:scale}). The effect of a finite temperature can be
incorporated in the present discussion by replacing the system size
$L$ by a phase coherence length $L_\Phi$ that diverges as the
temperature tends to zero.

From scaling theory, the following scenario for the temperature or
system size dependence emerges: on small length scales or at high
temperatures, classical Drude theory applies. At high magnetic fields,
the effects of Landau quantization become visible, when
$\pi^2E_T/E_B\approx1$ \cite{AFS82}. In GaAs and for $T=4.2$K this 
happens at a magnetic field of about 2T. At lower temperatures,
the Drude expression for $\sigma_{xx}^0$ is only valid up to about
$\omega_c\tau=1$ beyond which the SCBA result $\sigma_{xx}^0\lesssim
e^2/\pi h$ becomes appropriate. Localization effects leading to the
QHE become important when the system size and phase coherence length
exceed the localization length $\xi^0$. If $\sigma_0$ exceeds $e^2/h$,
this length scale very rapidly becomes larger than the phase coherence
length in present day experiments, provided $\omega_c\tau<1$.
However, around $\omega_c\tau=1$ the bare conductivity drops below
$e^2/h$ and $L_\Phi$ can exceed $\xi^0$ at low temperatures. In
particular, near integer filling factors the conductivity is very
small and the crossover length $\xi^0$ becomes small. The point
$\omega_c\tau\approx 1$ separates two regions with very different
temperature behaviors: at low fields, $\rho_{xx}$ increases slowly
with decreasing temperature, while at higher fields, $\rho_{xx}$
decreases, most 
strongly near integer filling factors, due to the onset of strong
localization on the quantum Hall plateau. At the crossover point the
resistivity will be only very weakly temperature dependent. Note
however, that this point does not correspond to a critical point in
the zero temperature phase diagram.  Up to the crossover point near
$\omega_c\tau=1$ deviations from Drude behavior are small so that near
the crossing point $\rho_{xx}=\rho_{xy}$.

We thus find that standard scaling theory predicts the essential
features of the experiments that have been interpreted as showing a
low-field QH-insulator transition: A magnetic field at which
$\rho_{xx}$ is temperature independent has been observed in various
experiments \cite{JJWH93Wea94,STC95,Sea97,LCSL98,Hea99}. This
``critical'' field separates an ``insulating'' low-field region with
weak temperature dependence from a metallic QH region with stronger
temperature dependence on the high field side. At this ``transition''
Hall and dissipative resistivity are approximately equal
\cite{Sea97,Hea99}. The value of the resistivities at this transition
is approximately $1/\sigma_0=1/en_c\mu$. The density dependence of
this values should thus follow the density dependence of the
zero-field mobility\cite{Sea97}.

It should be stressed, that the validity of this argument goes beyond
the validity of the employed approximations. The physical mechanism
responsible for the drastic change in the temperature dependence near
$\omega_c\tau=1$ is the suppression of the bare conductivity at high
fields due to the gaps in the density of states as a result of Landau
quantization. This leads to the strong field-dependence of the
crossover scale $\xi^0$. For a more quantitative agreement with
experiment, the bare conductivities should be evaluated in SCBA taking
into account Landau level mixing and higher order corrections should
be included in eq.~(\ref{eq:uni_cor}).

The question arises, under which conditions the non-monotonic
dependence of the Hall conductivity predicted by the levitation
scenario could be observed. Khmel'nitskii and Laughlin have argued
that the plateau transitions are given by the condition that the Drude
Hall conductivity $\sigma_{xy}^0$ equals half-integer multiples of
$e^2/h$. This implies a lower bound on $n_c\mu$ for the occurrence of
the QHE. The maximum value of $\sigma_{xy}^0$ is $\sigma_0/2$ at
$\omega_c\tau=1$. Thus, for $\sigma_0<e^2/h$ there are no plateau
transitions and hence no QHE. The reentrant plateau transitions occur
for $\omega_c\tau<1$, where Drude theory is the appropriate expression
for the bare conductivity. The minimum $\sigma_0$ for the occurrence
of the $n=2$ plateau is $3e^2/h$ and the minimum crossover length
$\xi^0$ at $\omega_c\tau=1$ is $l_c\exp((3\pi/2)^2)=4.4\cdot10^9l_c$,
a macroscopic quantity for magnetic fields in the Tesla range. The
zero temperature phase diagram with the levitating critical states at
low fields is thus of very little importance for experiments on the
QHE at low magnetic fields. Even though scaling theory predicts a very
different behavior at zero temperature, at all but exponentially low
temperatures it predicts a linear increase of the Hall resistivity up
to fields where $\omega_c\tau\approx1$ and the onset of monotonically
increasing quantum Hall plateaus beyond.

The system behaves differently, when only the $n=1$ plateau is
observable.  The bare conductivity at the low-field QH-insulator
transition can then be of the order of $e^2/2h$ and the crossover
scale $\xi^0$ can be microscopic. At this transition scaling behavior
should be observable.\cite{STC95,Sea97,Hea99bCea99}

From these consideration, we are led to conclude that the recent
experimental observation of a temperature independent resistivity at
low magnetic fields does, in fact, not contradict the scaling theory
of the QHE, but rather is an expected finite-temperature effect. We
further see that experiments on the low-field behavior of QH systems
reveal only very limited information on the zero-temperature quantum
phase transitions. In particular, they don't give much insight into
the nature of the insulator phase below the lowest QH transition. The
experiments can, however, help to improve our understanding of the
finite-size and finite-temperature effects associated with weak
localization.

The situation is quite different on the high magnetic field side.
Here, the SCBA applies as the starting point for the renormalization
of the conductivities \cite{WTP85} and the bare conductivity in the
lowest Landau level is less than $e^2/h$. Thus it is possible to reach
the asymptotic scaling regime, both in experiments and in numerical
simulations \cite{Huc95r}. The results for finite temperatures/system
sizes can reliably be extrapolated by finite-size scaling. However,
even here the nature of the insulating phase remains quite elusive
experimentally. In order to study the insulating phase, it is
necessary to go beyond the scaling region of the QH-insulator
transition. Numerically, it has been found that the Hall resistivity
remains quantized at $h/e^2$ throughout the region where scaling
behavior is observed\cite{SW98SW99}. At the high-field end of the
scaling region the longitudinal resistivity was found to be up to
$16h/e^2$ \cite{Sea98}, making accurate measurements of the much
smaller Hall resistivity difficult. At zero temperature the width of
the scaling region shrinks to zero. The experimentally observed
quantization of the Hall resistivity through the transition
\cite{Hea98aHea98b} is thus likely to be a confirmation of scaling
behavior and is no indication of the transport properties of the
insulating phase at zero temperature.

In conclusion, I have discussed the behavior of the quantum Hall
effect at low magnetic fields as expected from the scaling theory of
the QHE. The large localization length in a magnetic field in two
dimensions restricts the observability of the levitating critical
states to exponentially small temperatures and exponentially large
systems. At accessible temperatures and system sizes the Hall
resistivity will be a monotonically increasing function of magnetic
field. Near magnetic fields, such that $\omega_c\tau\approx1$, the
temperature dependence of the dissipative resistivity changes from
weakly increasing at low magnetic fields to decreasing at higher
magnetic fields, in accordance with recent experiments. At this
approximately temperature-independent point $\rho_{xx}$ and
$\rho_{xy}$ are of equal magnitude.

I acknowledge stimulating discussions with Z. Wang and X.C. Xie on the
quantum Hall insulator and the hospitality of the Institute for
Theoretical Physics at Santa Barbara. This work was performed within
the research program of the SFB 341 of the DFG and supported by the
NSF at ITP.

\enlargethispage*{1000pt}


\begin{thebibliography}{10}
\bibitem{scaling}
The term ``scaling theory'' is used to describe the global phase diagram
  \cite{KLZ92} and the renormalization of the conductivities
  \cite{Khm83Pru87}. For a discussion of the analytic structure of
  such a theory, see \cite{Zir99}.
\bibitem{greenwood}
The change of the conductivity with temperature due to the Fermi distribution
  is of no importance in the present context. High temperatures refers to the
  neglect of quantum interference contributions.
\end{thebibliography}
\end{document}